\documentclass[preprint]{JHEP3} 

\usepackage{epsfig,multicol}

\newcommand{\beq}{\begin{equation}} 
\newcommand{\eeq}{\end{equation}}
\newcommand{\bea}{\begin{eqnarray}}
\newcommand{\eea}{\end{eqnarray}}
\newcommand{\rf}[1]{(\ref{#1})}                                    %
\newcommand{\tr}{\mbox{Tr}\;}

\title{Rational three-spin string duals and non-anomalous finite size effects}

\author{L.\ Freyhult
                and C.\ Kristjansen\\
        NORDITA, Blegdamsvej 17, DK-2100 Copenhagen \O, Denmark\\
        E-mail: \email{freyhult@nordita.dk}, \email{kristjan@nbi.dk}}

\preprint{hep-th/0502122 \\ NORDITA-2005-12}     

\abstract{
We determine by a one line computation the
one-loop conformal dimension and the associated non-anomalous
finite size correction for all
operators dual
to 
spinning strings of rational type having three
angular momenta $(J_1,J_2,J_3)$
on $S^5$. Finite size 
corrections are conjectured to encode 
information about string sigma model loop corrections to the
spectrum
of type IIB superstrings on $AdS_5$ $\times$ $S^5$. 
We compare our
result to the zero-mode contribution to the 
leading quantum string correction derived for the
stable three-spin string with two out of the three spin labels identical
and observe agreement. 
As a side result we clarify the relation between the Bethe root
description of three-spin strings of the type $(J,J',J')$ with
respectively $J>J'$ and $J<J'$.
}


\begin{document} 


\section{Introduction}
Semi-classical analysis of strings propagating on $AdS_5\times S^5$ ,
see~\cite{Tseytlin:2003ii} 
for a review, has provided us with very concrete realizations
of the AdS/CFT duality~\cite{Maldacena:1997re}. The prime example is the
matching of the energies of free strings with conformal dimensions of 
operators of planar ${\cal N}=4$ SYM. Strings amenable to semi-classical
analysis are strings carrying large quantum numbers. For strings with
several large angular momenta on AdS$_5\times S^5$, at least one
of them lying on S$^5$, 
the semi-classical analysis is particularly
clean~\cite{Frolov:2003qc,Frolov:2003tu}. The classical
string energy organizes into a power series in $\frac{\lambda}{J^2}$ where
$J$ is the sum of angular momenta and $\lambda$ is the squared string
tension which via the AdS/CFT dictionary is mapped onto
the 't Hooft coupling
of the dual gauge theory. Furthermore, string sigma model loop corrections are suppressed
by powers of $\frac{1}{J}$ 
compared to the classical energy. More precisely, the
combined expansion of the semi-classical energy in $\frac{\lambda}{J^2}$
and $J$ takes the form~\cite{Frolov:2003qc,Frolov:2003tu}
\beq
E=J\left\{1+\frac{\lambda}{J^2}
\left(E_1^{(0)}+\frac{1}{J}E_1^{(1)}+\ldots\right)
+\left(\frac{\lambda}{J^2}\right)^2
\left(E_2^{(0)}+\frac{1}{J}E_2^{(1)}+\ldots\right)+\ldots\right\},
\label{exp0}
\eeq
where the classical energy is given by
\beq
E_{cl}=J\left\{1+\frac{\lambda}{J^2}E_1^{(0)}+
\left(\frac{\lambda}{J^2}\right)^2E_2^{(0)}+\ldots\right\}.
\label{expansion}
\eeq
Based on the AdS/CFT conjecture one expects a similar reorganization of the
perturbative series for the conformal dimension of the dual operator to be
possible. In gauge theory language angular momenta
are simply representation
labels  and $J$ is the sum of such labels.
It is thus tempting to attempt a comparison with string theory,
calculating conformal dimensions by first doing a perturbative expansion and 
subsequently taking the limit \mbox{$J\rightarrow\infty$} 
with $\frac{\lambda}{J^2}$
fixed. This approach was very successful at the one-loop level where
the dilatation operator of the gauge theory could be proved identical
to the Hamiltonian of an integrable spin 
chain~\cite{Minahan:2002ve,Beisert:2003jj,Beisert:2003yb} and diagonalization
could be carried out in a number of specific cases~
using Bethe equation techniques~\cite{Beisert:2003xu,
Beisert:2003ea,Engquist:2003rn,Kristjansen:2004ei,Kristjansen:2004za}.
It also became possible to prove the equivalence of the semi-classical
treatment of the string and the perturbative treatment of the gauge theory
at order $\lambda$ at a more general level, not referring to particular
solutions~\cite{Kruczenski:2003gt,Kazakov:2004qf,Hernandez:2004uw,
Kruczenski:2004kw,Stefanski:2004cw,Hernandez:2004kr,Kazakov:2004nh,
Beisert:2004ag,Schafer-Nameki:2004ik}. 
Encouraged by this success there are several paths one can take. 

One can try
to incorporate higher order terms in the perturbative analysis of the
gauge theory. This should allow one to reproduce more terms in the
expansion~\rf{expansion} of the classical string energy. As the dilatation
operator of the gauge theory is known to higher loop orders in certain
sub-sectors~\cite{Beisert:2003tq,Beisert:2003ys}, see
also~\cite{Staudacher:2004tk}, this line of investigation is indeed
possible and was initiated in~\cite{Serban:2004jf}. The matching 
with the semi-classical string analysis worked successfully at two
loops but at three loops a discrepancy was observed~\cite{Serban:2004jf},
see also~\cite{Minahan:2004ds}.  
An explanation of
this discrepancy as nothing but a manifestation of the strong/weak
coupling nature of the AdS/CFT correspondence has been put
forward~\cite{Serban:2004jf,Beisert:2004hm}.

 Another path one can take to further investigate the relation between
${\cal N}=4$ SYM and semi-classical strings is to attempt to go beyond
the planar approximation of the gauge theory. In string theory
language this implies taking into account string interactions. While
an interesting suggestion for how to deal semi-classically with the
splitting of certain strings propagating on $AdS_5\times S^5$
exists~\cite{Peeters:2004pt}, unfortunately not much progress has been
made in the development of the necessary calculational techniques on
the gauge theory side.

Finally, one could attempt to study the gauge theory counterpart of
string sigma model
loop corrections and this is the line of investigation we shall follow
here.
The one-loop string correction has been known for some time for one
particular string configuration, namely a circular string rotating on 
$S^5$ and carrying centre of mass angular momentum $J_1$ and two equal
angular momenta $J_2=J_3$ with respect to two orthogonal planes in
$S^5$. The string is stable for $J_2$ small enough and in the region
of stability the one loop correction has been
calculated~\cite{Frolov:2003qc,Frolov:2003tu,Frolov:2004bh}. Actually,
the calculation of this one-loop correction provided the rationale for
all subsequent semi-classical analysis of strings propagating in
$AdS_5\times S^5$. On the gauge theory side, assuming the dilatation
operator to be given by the Hamiltonian of an integrable spin chain,
quantum corrections to the classical string energy translates into
finite size corrections to energy eigenvalues of the infinite chain.
In the present paper we shall determine the 
non-anomalous part of the 
finite size correction to
the conformal dimension (i.e.\ spin chain eigen energy) for the
operator
dual to the above mentioned three-spin string. Actually, we shall do
much more than that. We shall determine the non-anomalous part of the
finite size correction for
all operators dual to so-called rational strings carrying three
angular momenta $(J_1,J_2,J_3)$ on $S^5$. We shall furthermore
compare 
the result obtained for the dual of the stable
three-spin string above to the zero-mode contribution to the
string quantum correction and find agreement. 
So far non-anomalous 
finite size contributions 
were only determined for the
simpler case of the dual of an unstable string with two spins on
$S^5$~\cite{Lubcke:2004dg} and the case of a stable string with one
spin on $S^5$ and one on $AdS_5$~\cite{Kazakov:2004nh}. Very recently
the one-loop string correction for the latter string was determined
and agreement between zero-mode contribution and non-anomalous finite
size effects in gauge theory likewise found~\cite{Park:2005ji}.

In section~\ref{spinningstrings} we shall
review the properties of strings  rotating on $S^5$ and their gauge
theory duals. After that in section~\ref{finitesize} we calculate
the non-anomalous part of the
finite size correction to the conformal dimensions of operators
dual to rational
three-spin strings and compare in section~\ref{comparison} to the known
result for the stable three-spin string with two out of the three
angular momenta coinciding. Subsequently, in section~\ref{highercharges} 
we explain
how one can easily obtain also all higher charges of the integrable spin
chain in question and the associated non-anomalous 
finite size corrections. Finally
in section~\ref{particular} 
we discuss some particular points  of 
our solution. 
Section~\ref{conclusion} contains our conclusions.

After the completion of this manuscript we
received~\cite{Beisert:2005mq} where it was demonstrated that for
a two-spin string with one large angular momentum on $S^5$ and one on $AdS_5$
the finite size corrections have a non-anomalous part which matches
the contribution to the string sigma model loop correction coming from
non-zero modes.

\section{Strings spinning on $S^5$ \label{spinningstrings}
and their gauge theory duals}

One class of strings spinning on $S^5$ is particularly simple, namely
the class of circular, rigid strings carrying centre of mass angular
momentum
$J_1$ and two equal angular momenta $J_2=J_3$ with respect to two
orthogonal
planes in $S^5$. Denoting by $J=J_1+J_2+J_3$ the total angular
momentum
and by $k$ the winding number the first terms in the
expansion~\rf{expansion}
of the classical energy of such strings take the
form~\cite{Frolov:2003qc,Frolov:2003tu}
\beq
E_{cl}=J+\lambda\, k^2 \frac{J_2}{J^2},
\eeq 
The above strings form a sub-class of the so-called rational
three-spin
strings~\cite{Arutyunov:2003za}. A general  three-spin string of rational
type
with angular momenta $(J_1,J_2,J_3)$ is characterized by the angular
momenta being related by
\beq
m_1J_1+m_2J_2+m_3J_3=0,
\eeq
with $m_1$, $m_2$ and $m_3$ integer and by the first terms in the
large-$J$ expansion of the classical energy being given by
\beq
E_{cl}=J+\frac{\lambda}{2J} \sum_{i=1}^3 m_i^2 \,\frac{J_i}{J}.
\eeq
The simpler case with $J_2=J_3$ is recovered for $m_1=0$, $m_2=-m_3=k$.
In addition to the rational type, three-spin strings come in an
elliptic
and in a hyper-elliptic version, the terminology referring to the type
of functions needed to parametrize the classical string sigma model
solution~\cite{Arutyunov:2003uj}. 
Only for the simplest rational three-spin string with
angular
momenta~$(J_1,J_2,J_3)=(J_1,J_2,J_2)$ a study of one-loop sigma model
corrections has been
possible~\cite{Frolov:2003qc,Frolov:2003tu,Frolov:2004bh}.

At one loop order in $\lambda$ 
rigid strings spinning on $S^5$ with angular momenta
$(J_1,J_2,J_3)$ are dual to operators of the type $\tr(X^{J_1} Y^{J_2}
Z^{J_3}
+perm's)$, where $X$, $Y$ and $Z$ are the three complex scalars
of ${\cal N}=4$ SYM, and the conformal dimension of such operators  can
be found by diagonalizing the Hamiltonian of the ferromagnetic
$SU(3)\subset SO(6)$ spin chain of length $J=J_1+J_2+J_3$. The
diagonalization is carried out by solving a set of algebraic equations
for the Bethe roots~\cite{Bethe:1931hc}. 
Representations of $SO(6)$ can be labelled
by three highest weight labels $(J_1,J_2,J_3)$ with $J_1\geq J_2\geq
J_3$
 or equivalently by three Dynkin indices
$[d_1,d_2,d_3]=[J_2-J_3,J_1-J_2,J_2+J_3]$.
For the $SU(3)$ spin chain there are two types of Bethe 
roots~$\{u_{1,j}\}_{j=1}^{n_1}$ and~$\{u_{2,j}\}_{j=1}^{n_2}$ and the
number of these, $n_1$ and $n_2$, determine the representation 
as follows
\beq
[d_1,d_2,d_3]=
[n_1-2n_2,J-2n_1+n_2,n_1],
\eeq
which implies\footnote{We use the notation $(J_1,J_2,J_3)$ both for
  representation labels on the gauge theory side and for angular
  momenta
on the string theory side. When referring to gauge theory quantities
it is always assumed that $J_1\geq J_2\geq J_3$ whereas in general no
ordering is assumed when referring to string theory quantities.} 
\beq
(J_1,J_2,J_3)=(J-n_1,n_1-n_2,n_2).
\label{amomenta}
\eeq
 The Bethe equations read
\bea
\left(\frac{u_{1,j}+i/2}{u_{1,j}-i/2}\right)^J
&=&\prod_{k\neq j}^{n_1} 
\frac{u_{1,j}-u_{1,k}+i}{u_{1,j}-u_{1,k}-i}
\prod_{k=1}^{n_2} \frac{u_{1,j}-u_{2,k}-i/2}{u_{1,j}-u_{2,k}+i/2},
\label{Bethe1} \\
1&=&\prod_{k\neq j}^{n_2} 
\frac{u_{2,j}-u_{2,k}+i}{u_{2,j}-u_{2,k}-i}
\prod_{k=1}^{n_1} \frac{u_{2,j}-u_{1,k}-i/2}{u_{2,j}-u_{1,k}+i/2}.
\label{Bethe2}
\eea
The spin chain eigenstates sought for must reflect the characteristics
of a trace implying first that the spin chain must be considered
periodic
and secondly that a zero momentum condition must be imposed, i.e.\
\beq
1=\prod_{j=1}^{n_1}\left(\frac{u_{1,j}+i/2}{u_{1,j}-i/2}\right).
\label{cyclicity}
\eeq
A solution of the Bethe equations supplemented by~\rf{cyclicity} gives
rise to an eigen operator with one-loop conformal dimension
\beq
\label{gamma}
\gamma_1=\frac{J}{8\pi^2}\sum_{j=1}^{n_1}\frac{1}{(u_{1,j})^2+1/4},
\eeq
where in analogy with eqn.~\rf{exp0} we have for the conformal dimension
\beq
\gamma=J\left\{1+\frac{\lambda}{J^2}\gamma_1
+\left(\frac{\lambda}{J^2}\right)^2 \gamma_2+\ldots \right\}.
\eeq
Rescaling all roots by a factor of $J$, i.e.\ setting $u_{i,j}=J\,
q_{i,j}$,
taking the logarithm  of the equations~\rf{Bethe1} and \rf{Bethe2}, 
 and expanding for large $J$
one finds 
\begin{eqnarray}
\label{Bethe1a}\frac{1}{q_{1,k}}+2\pi n_k&=
&\frac{2}{J}\sum_{l\neq k}^{n_1}\frac{1}{q_{1,k}-q_{1,l}}-\frac{1}{J}
\sum_{l=1}^{n_2}\frac{1}{q_{1,k}-q_{2,l}},\\
\label{Bethe2a}2\pi m_k&=&\frac{2}{J}\sum_{l\neq
  k}^{n_2}\frac{1}{q_{2,k}-q_{2,l}}-\frac{1}{J}\sum_{l=1}^{n_1}\frac{1}{q_{2,k}-q_{1,l}}.
\end{eqnarray}
Solving these equations to leading and next to leading order in $\frac{1}{J}$
gives the one-loop 
conformal dimension and the associated non-anomalous part of its
$\frac{1}{J}$ correction. An additional anomalous piece originating from
nearby Bethe roots with $q_j-q_k\sim {\cal O}(\frac{1}{J})$ is not taken into
account by these formulas~\cite{Beisert:2005mq}.
By the same procedure we get for the zero momentum condition 
\beq
\label{cyclicitya}-2\pi p=\frac{1}{J}\sum_{l=1}^{n_1}\frac{1}{q_{1,l}}.
\eeq
Here $n_k$, $m_k$ and $p$ are all integers and reflect the ambiguity
in the choice of branch for the logarithm.
Finally, after the rescaling the expression for $\gamma_1$ takes the form
\beq
\label{gammaa}
\gamma_1=\frac{1}{8\pi^2 J}\sum_{j=1}^{n_1}\frac{1}{(q_{1,j})^2}.
\eeq

\section{Rational three-spin strings \label{finitesize} }

In the following we shall consider the case of rational three spin
strings. These can be reached by choosing
\bea
n_k &=& n, \hspace{0.5cm} \forall \,\,\, k\in \{1,\ldots,n_1\}, \label{nk}\\
m_k &=& m, \hspace{0.5cm} \forall \,\,\,k \in \{1,\ldots,n_2\} \label{mk}.
\eea
Then our Bethe equations trivially reduce to
\begin{eqnarray}
\label{Bethe1b}
\frac{1}{q_{1,k}}+2\pi n&=
&\frac{2}{J}\sum_{l\neq k}^{n_1}\frac{1}{q_{1,k}-q_{1,l}}-\frac{1}{J}
\sum_{l=1}^{n_2}
\frac{1}{q_{1,k}-q_{2,l}},\\
\label{Bethe2b}
2\pi m&=&\frac{2}{J}\sum_{l\neq
  k}^{n_2}\frac{1}{q_{2,k}-q_{2,l}}-\frac{1}{J}\sum_{l=1}^{n_1}\frac{1}{q_{2,k}-q_{1,l}}.
\end{eqnarray}
Similar equations appear in the study of the matrix model of ${\cal N}=2$
supersymmetric type $U(N)\times U(N)$ gauge theory~\cite{Hailu:2004dq}.
In order to expose the similarity of the following analysis with the 
loop equation approach to matrix models we shall introduce the notation
\beq
V_1'(q)\equiv\frac{1}{q}+2\pi n, \hspace{0.7cm}V_2'(q)\equiv 2\pi m.
\eeq
In addition, let us introduce the resolvents by
\beq
G_1(q)=\frac{1}{J} \sum_{j=1}^{n_1}\frac{1}{q-q_{1,j}},\hspace{0.5cm}
G_2(q)=\frac{1}{J} \sum_{j=1}^{n_2}\frac{1}{q-q_{2,j}},
\label{resolvents}
\eeq
and the filling fractions
\beq
\alpha=\frac{n_1}{J}, \hspace{0.5cm}\beta=\frac{n_2}{J}.
\eeq
Then we have 
\bea
G_1(q)&\sim& \frac{\alpha}{q}\hspace{0.5cm}\mbox{as}\hspace{0.5cm} q\rightarrow
\infty, \label{G1asymptotic}\\
G_2(q)&\sim& \frac{\beta}{q}\hspace{0.5cm}\mbox{as}\hspace{0.5cm} q\rightarrow
\infty, \label{G2asymptotic}
\eea
as well as
\beq
G_1(0)=2\pi p, \label{G10}
\eeq
and
\beq
\gamma_1=-\frac{1}{8\pi^2 }G_1'(0).
\label{bc}
\eeq
To determine $\gamma_1$ we first multiply eqn.~\rf{Bethe1b} 
by
$\frac{1}{J^2}\sum_{m=1}^{n_2}\frac{1}{q_{2,m}-q_{1,k}}$ and sum
over $k$. Next, we multiply eqn.~\rf{Bethe2b}
by $\frac{1}{J^2}\sum_{m=1}^{n_1}\frac{1}{q_{1,m}-q_{2,k}}$ and sum over
$k$ and finally we subtract the resulting two equations. This gives
us the following relation
\beq
-G_1'(0)\left(1-\frac{1}{J}\right)-4\pi n G_1(0)
+(2\pi n)^2 \alpha+ (G_1(0))^2-(2\pi m)^2\beta =0.
\label{Gprime}
\eeq
Furthermore, simply summing over $k$ in eqns.~\rf{Bethe1b} and~\rf{Bethe2b}
leads to
\beq
p=n\alpha+m\beta. \label{momentum}
\eeq
Bearing in mind the relations~\rf{G10} and~\rf{bc} 
and writing
\beq
\gamma_1=\gamma_1^{(0)}+\frac{1}{J}\gamma_1^{(1)},
\eeq
we see that 
eqn.~\rf{Gprime} allows us
to compute the conformal dimension as well as its leading 
non-anomalous
$\frac{1}{J}$
correction.
The leading order result for the conformal dimension reads
\beq\label{gamma0}
\gamma_1^{(0)}=\frac{1}{2}
\left(n^2\alpha(1-\alpha)+2\beta m n (1-\alpha)+m^2\beta(1-\beta)
\right).
\eeq
Translating from filling fractions $(\alpha,\beta)$ to angular
momenta $(J_1,J_2,J_3)$, cf.\ eqn.~\rf{amomenta} one sees that eqn.~\rf{gamma0}
 is exactly the expression
characteristic of the general rational three-spin string upon the
identification
\beq
m_1=p,\hspace{0.5cm}m_2=p-n,\hspace{0.5cm}m_3=p-m-n.
\eeq
The formula~\rf{gamma0} and the
 accompanying constraint~\rf{momentum} are invariant
under the transformations
\bea
1-\alpha\leftrightarrow \alpha-\beta, \hspace{0.7cm}& n\rightarrow -n, &
\hspace{0.7cm} m\rightarrow m+n,\hspace{0.7cm}p\rightarrow p-n, 
\label{transf1}\\
\alpha\leftrightarrow 1-\beta, \hspace{0.7cm} & m\leftrightarrow -n, &
\hspace{0.7cm} p\rightarrow p-m-n,
\label{transf2}\\
\alpha-\beta\leftrightarrow \beta, \hspace{0.7cm}& m\rightarrow -m, &
\hspace{0.7cm} m\rightarrow m+n, \hspace{0.7cm} p\rightarrow p.
\label{transf3}
\eea
These invariances are the generalizations of the invariance of the
spin-1/2 Heisenberg chain under the interchange of spin up and spin
down.
The latter is recovered from relation~\rf{transf1} when $\beta=0$, $m=0$.

Including the $\frac{1}{J}$ correction simply gives
$\gamma_1^{(1)}=\gamma_1^{(0)}$ or
\beq
\gamma_1=\gamma_1^{(0)}\left(1+\frac{1}{J}\right),
\eeq
with $\gamma_1^{(0)}$ given above.

So far the Bethe root description of operators dual to rational three
spin strings was given only for the case of two coinciding spin 
labels~\cite{Engquist:2003rn}. Such operators are of course included
in the present analysis. 
Strings with angular momenta
$(J_1,J_2,J_2)$ where $J_1>J_2$ and winding number $k$ we can reach by choosing
\beq
m=-2n,\hspace{0.7cm} p=0, 
\hspace{0.7cm} \mbox{and}\hspace{0.7cm} m=k.
\label{JJ'J'}
\eeq
The momentum constraint~\rf{momentum} here implies 
\beq
\alpha=2\beta, \label{JJ'J'2}
\eeq
and therefore $(J_1,J_2,J_2)=J(1-\alpha,\frac{\alpha}{2},\frac{\alpha}{2})$
where it is understood that
$\alpha\in [0,\frac{2}{3}]$.
Strings 
with angular momenta $(J_1,J_1,J_3)$ where $J_1>J_3$ 
and winding number $k$
are reached 
by setting 
\beq
\label{JJJ'}
n=-2m, \hspace{0.7cm} p=-m,\hspace{0.7cm}
\mbox{and}\hspace{0.7cm} n=k.
\eeq
This choice of parameters implies 
\beq
2\alpha=\beta+1, \label{JJJ'2} 
\eeq
and thus $(J_1,J_1,J_3)=J(1-\alpha,1-\alpha,2\alpha-1)$
where it is understood
that $\alpha\in [\frac{1}{2},\frac{2}{3}]$.
In both of these cases we have
\beq
\gamma_1^{(0)}=\gamma_1^{(1)}=k^2\frac{J_2}{J}. \label{gamma11}
\eeq
We notice that the winding number of the string
in the former case has to be identified with the mode number $n$ while
in the latter case it has to be identified with the mode number $m$,
cf.\ eqns.~\rf{Bethe1b} and~\rf{Bethe2b}.
Strictly speaking, the expression~\rf{Gprime} only constitutes  a 
necessary condition for the Bethe equations to be fulfilled. However,
we shall explain in section~\ref{highercharges} how we can determine
all the higher charges of the spin chain as well, thus obtaining the
full solution of the Bethe equations (including the relation~\rf{Gprime}).

Finally, let us mention that by setting 
\beq
m=0, \hspace{0.7cm} \beta =0,
\eeq
and thus $p=n\alpha$, cf.\ eqn.~\rf{momentum},
we recover the 
result for the conformal dimension and the associated 
non-anomalous part of the finite 
size correction obtained in~\cite{Lubcke:2004dg} for the case of a
rational two-spin string with 
angular momentum assignment
$(J_1,J_2,0)=J(1-\alpha,\alpha,0)$ and winding number $n$.

\section{Comparison to string theory \label{comparison} }

As already mentioned one sub-class of rational three-spin strings is
particularly manageable, namely the sub-class of strings having two out
of the three angular momenta coinciding, i.e. $(J_1,J_2,J_3)=(J_1,J_2,J_2)$.
For such strings the Lagrangian for the quadratic fluctuations around 
the classical solution involves only constant coefficients. This means
that a stability analysis can immediately be carried out and in case of 
stability the one-loop correction to the energy can 
found as the sum of the characteristic bosonic and fermionic frequencies. 
The stability analysis 
gives a clear answer: the string is stable provided
a parameter $q\approx\frac{2J_2}{J}$ fulfils the
relation~\cite{Frolov:2003qc,Frolov:2003tu,Frolov:2004bh}
\beq
q<q_c=1-(1-\frac{1}{2k})^2,
\eeq
where $k$ is the winding number of the string. Translating to angular
momenta this (approximately) reads
\beq
J_2<\frac{1-(1-\frac{1}{2k})^2}{2(1-\frac{1}{2k})^2} J_1.
\eeq
In particular, we see that the two-spin version of this string corresponding
to $J_1=0$ is always unstable. For $J_1\neq0$ there is always a certain
stability region and it appears that this stability region at least for 
$k\geq2$ lies entirely within the class of strings that we can reach 
with the parameter choice given in eqn.~\rf{JJ'J'}. (As it will
appear from section~\ref{particular} we also have access to the
$k=1$ case with the same parameter choice.) 
The actual computation of the one-loop string correction, i.e.\ the
summing over bosonic and fermionic frequencies, 
is rather
involved and can only be  carried out numerically. 
We note that the bosonic frequencies can be reproduced in the gauge
theory language by studying Bethe root
fluctuations~\cite{Beisert:2003xu,Freyhult:2004iq}.
It is, however, hard to see how the fermionic frequencies would be
encoded
in the Bethe root picture. 
The zero-mode contribution to the one-loop string correction is found
to be~\cite{Frolov:2003qc,Frolov:2003tu,Frolov:2004bh}
\beq
E_{1,zero}^{(1)}=k^2 \frac{J_2}{J}.
\eeq
This is identical to our result for the non-anomalous part of the 
$\frac{1}{J}$ correction $\gamma_1^{(1)}$, cf.\ eqn.~\rf{gamma11}.

\section{Higher charges \label{highercharges} }
Above we have derived the one-loop 
conformal dimension including the non-anomalous finite size corrections
for all rational three-spin strings by essentially 
a one line computation. In a similarly
simple fashion one can derive expressions for the higher charges 
of the spin chain and the associated non-anomalous
finite size corrections. The most efficient way to do
so, however, is to determine in one step the 
resolvent $G_1(q)$ as it is well-known that this function 
acts as a generator of the higher 
charges~\cite{Arutyunov:2003rg,Engquist:2003rn}, see 
also~\cite{Engquist:2004bx}. It is
straightforward to derive algebraic equations which determine 
$G_1(q)$ as well as $G_2(q)$. 
More precisely, we can derive a quadratic and a cubic equation which
when combined determine the two resolvents. To derive the quadratic
equation  we first multiply  eqn.~\rf{Bethe1b} by
$\frac{1}{J}\frac{1}{q-q_{1,k}}$ and sum over $k$, subsequently
multiply
eqn.~\rf{Bethe2b} by $\frac{1}{J}\frac{1}{q-q_{2,k}}$ and sum over $k$
and finally add the two resulting equations. This gives
\bea
&&G_1^2(q)+G_2^2(q)-G_1(q)G_2(q)\nonumber \\
&&-V_1'(q)G_1(q)-V_2'(q)G_2(q)+
\frac{1}{q}G_1(0)+\frac{1}{J}\left(G_1'(q)+G_2'(q)\right)=0.
\label{quadratic}
\eea
To derive the cubic equation we take a similar strategy. First, we
multiply
the relation~\rf{Bethe1b} by
$\frac{1}{J^2}\sum_{m=1}^{n_2}\frac{1}{q_{2,m}-q_{1,k}}\frac{1}{q-q_{1,k}}$
and sum over $k$, subsequently multiply the 
relation~\rf{Bethe2b} by
$\frac{1}{J^2}\sum_{m=1}^{n_1}\frac{1}{q_{1,m}-q_{2,k}}\frac{1}{q-q_{2,k}}$
and sum over $k$
and finally we subtract the two resulting equations. The result
of these manipulations reads
\bea
&&G_2^2(q)G_1(q)-G_1^2(q)G_2(q) \nonumber \\
&&+V_1'(q)\left(G_1^2(q)+\frac{1}{J}G_1'(q)\right)
 -V_2'(q)\left(G_2^2(q)+\frac{1}{J}G_2'(q)\right)
\nonumber \\
&& -(V_1'(q))^2G_1(q)+(V_2'(q))^2G_2(q) \nonumber\\
&&+
\frac{1}{J}G_2'(q)G_1(q)-\frac{2}{J}G_1(q)G_1'(q)
-\frac{1}{J^2}G_1''(q) -\frac{1}{Jq^2}G_1(q)\nonumber\\
&& +\frac{1}{q^2}G_1(0)
+\frac{4 \pi n}{q}G_1(0)+\frac{1}{q}G_1'(0)-\frac{1}{q}G_1^2(0)
-\frac{1}{Jq}G_1'(0)+\frac{1}{Jq^2}G_1(0)=0.\label{cubic}
\eea
A similar set of equations has been derived for the previously mentioned matrix
model of ${\cal N}=1$ supersymmetric A$_2$ type $U(N)\times U(N)$ gauge
theory~\cite{Hailu:2004dq}. These equations can easily be solved to leading
and next to leading order in $\frac{1}{J}$. In particular, we can immediately discard the term
involving $G_1''(q)$. Next, we have that 
all terms involving $G_1'(q)$ and 
and $G_2'(q)$ only appear to next to leading order in $\frac{1}{J}$.
Thus in the large-$J$ limit the equations~\rf{quadratic} and~\rf{cubic}
become equations involving only the resolvents themselves. Here one should
remember that
$G_1(0)$ as well as $G_1'(0)$
are known quantities, cf. eqns.~\rf{G10} and~\rf{Gprime}.
The quantity $G_1(0)$ is determined by the zero momentum condition,
cf.\ eqn.~\rf{cyclicitya} and $G_1'(0)$ is determined
by the asymptotic behaviour of the resolvents as $q\rightarrow \infty$. 
Actually, the relation~\rf{Gprime} is nothing but the leading part
of eqn.~\rf{cubic} as $q\rightarrow\infty$. 
We note that the equation~\rf{quadratic} in the large-$J$ 
limit has a striking similarity with the loop
equation of the $O(-1)$ matrix model and accordingly one can conveniently
split the resolvents into their regular and singular parts as 
follows~\cite{Kostov:1988fy}
\bea
G_1(q)&=&\frac{1}{3}\left(2V_1'(q)+V_2'(q)\right) +g_1(q), \\
G_2(q)&=&\frac{1}{3}\left(2V_2'(q)+V_1'(q)\right) +g_2(q),
\eea
where $g_1(q)$ and $g_2(q)$ constitute the singular parts. Inserting
this in  
eqns.~\rf{quadratic} and~\rf{cubic} we find to leading order in
$\frac{1}{J}$.
\bea
g_1^2(q)+g_2^2(q)-g_1(q)g_2(q)&=&r(q),\\
g_1^2(q)g_2(q)-g_2^2(q) g_1(q)&=&s(q),
\eea
where $r(q)$ and $s(q)$ are regular functions which read
\bea
r(q)&=&\frac{1}{3}\left((V_1'(q))^2+(V_2'(q))^2+V_1'(q)V_2'(q)\right)
-\frac{G_1(0)}{q},\label{rq}\\
s(q)&=&\frac{2}{3}(V_1'(q)-V_2'(q))r(q)+\frac{8}{27}((V_2'(q))^3-(V_1'(q))^3)
+\frac{1}{9}V_1'(q)V_2'(q)\left(V_2'(q)-V_1'(q)\right)\nonumber  \\
&&+\left(\frac{1}{q^2}+\frac{4\pi n}{q}\right)G_1(0)+\frac{1}{q}G_1'(0)
-\frac{1}{q}G_1^2(0). \label{sq}
\eea
or more explicitly
\bea
r(q)&=&\frac{1}{3q^2}-
\frac{2\pi}{q}\left\{n\left(\alpha-\frac{2}{3}\right)+ 
m\left(\beta-\frac{1}{3}\right)\right\}+
\frac{(2\pi)^2}{3}\left(m^2+n^2+mn\right), 
\label{rqexp}\\
s(q)&=&-\frac{2}{27q^3}+
\frac{2\pi}{3q^2}\left\{n\left(\alpha-\frac{2}{3}\right)
+m\left(\beta-\frac{1}{3}\right)\right\} \nonumber \\
&&+\frac{(2\pi)^2}{3q}\left\{ n^2\left(\alpha-\frac{2}{3}\right)- 
m^2\left(\beta-\frac{1}{3}\right)+
2mn\left(\alpha-\beta-\frac{1}{3}\right)\right\}\nonumber \\
&&+\left(\frac{2\pi}{3}\right)^3
\left( m- n\right)\left(2 n^2+2m^2+5 m n\right).
\label{sqexp}
\eea
The singular parts of the resolvents thus fulfill the following
cubic equations
\bea
(g_1(q))^3-r(q)g_1(q)&=&s(q),\label{g1cubic} \\
 (g_2(q))^3-r(q)g_2(q)&=&-s(q),\label{g2cubic}
\eea
which can readily be solved. 
We notice that the functions $r(q)$ and $s(q)$ are again invariant
under the transformations~\rf{transf1},~\rf{transf2} and~\rf{transf3}. 
Thus all higher charges possess these invariances as well.

Let us write down the explicit expressions for $r(q)$ and~$s(q)$ for
the case where two out of the three spin labels coincide. 
For the choice of parameters given in eqn.~\rf{JJ'J'} and~\rf{JJ'J'2}
i.e.\ for spin assignment
$(J_1,J_2,J_2)=J(1-\alpha,\frac{\alpha}{2},\frac{\alpha}{2})$ with
$\alpha\in[0,\frac{2}{3}]$ and thus
$J_1>J_2$   we get
\bea
r(q)&=&(2 \pi k)^2+\frac{1}{3q^2}, \label{r1}\\
s(q)&=&-\frac{2}{27}\frac{1}{q^3}-(2\pi k)^2\left(\alpha-\frac{2}{3}\right)
\frac{1}{q}.
\label{s1}
\eea
For the choice of parameters given in~\rf{JJJ'} and~\rf{JJJ'2}, i.e.
for spin assignment \\
$(J_1,J_1,J_3)=J(1-\alpha,1-\alpha,2\alpha-1)$
with $\alpha\in[\frac{1}{2},\frac{2}{3}]$ and thus $J_1>J_3$ 
we find
\bea
r(q)&=&(2 \pi k)^2+\frac{1}{3q^2}, \label{r2}\\
s(q)&=&-\frac{2}{27}\frac{1}{q^3}-
(2\pi k)^2\left(2(1-\alpha)-\frac{2}{3}\right)
\frac{1}{q}.
\label{s2}
\eea
We notice that when expressed in terms of the 
doubly degenerate
spin label, i.e.\ respectively $\frac{\alpha}{2}$ and $(1-\alpha)$ the
two sets of equations~\rf{r1}, \rf{s1} and~\rf{r2}, \rf{s2} coincide.
It is easy to see that the same is then the case for all the higher charges of the two string
duals.
A consequence of this is also that the expressions~\rf{r1} and~\rf{s1}
remain valid if analytically continued to the formally forbidden
parameter region $\alpha\in[\frac{2}{3},1]$ (and similarly for \rf{r2}
and~\rf{s2}).

The operators dual to rational three-spin strings with two out of the
three spin labels coinciding were earlier studied 
in reference~\cite{Engquist:2003rn}. Here the starting point was for 
both situations a seemingly more specialized assumption
about the root configuration. For spin assignment $(J_1,J_1,J_2)$,
$J_1>J_2$
the roots $\{q_{1,j}\}_{j=1}^{n_1}$ were assumed to 
be living on two distinct contours, for which the associated mode numbers
were respectively $k$ and $-k$, and which were each others mirror images 
with respect to the imaginary axis.
Furthermore, the roots $\{q_{2,j}\}_{j=1}^{n_2}$ 
were supposed to spread out over
the entire imaginary axis. 
This meant that  the rational three-spin string in question was effectively 
reached as a limiting
case of an elliptic three-spin string~\cite{Kristjansen:2004ei}.
From the double contour assumption an equation for the leading $J$ contribution
to the singular part of the resolvent for the roots $\{q_{1,j}\}_{j=1}^{n_1}$
was derived. This equation is
{\it exactly} identical to~\rf{g1cubic}, \rf{r1} and~\rf{s1}.
Thus, all  charges of the Bethe state of~\cite{Engquist:2003rn}
agree with those of ours to leading order in $\frac{1}{J}$. The two
states are therefore indistinguishable as descriptions of the classical 
three-spin string.
The case $(J_1,J_1,J_3)$ with $J_1>J_3$ was treated in
reference~\cite{Engquist:2003rn}  starting from a root configuration
which
included a so-called condensate. 
Also in this case  the rational
three-spin string in question was effectively obtained
as a limiting case of an elliptic three-spin string~\cite{Kristjansen:2004za}.
The condensate assumption did not immediately lead to a simple
equation
for the resolvent. It was nevertheless suggested that the condensate
solution was the analytical continuation of the double contour
solution
to the formally forbidden region of parameter space. Here, we have
obtained
a simple equation for the resolvent and we have seen explicitly how
the
analytical continuation works. 
While it
is well-understood in terms of algebraic geometry 
that the assumption of a condensate is redundant~\cite{Kazakov:2004qf}, 
it is less
clear why the double contour assumption of~\cite{Engquist:2003rn} for
the case $(J_1,J_2,J_2)$ with $J_1>J_2$ is equivalent to the treatment
given here.

We note that having obtained the leading order contributions
to the resolvents it is straightforward to determine the 
non-anomalous part of the 
next to leading
order ones as well. This simply requires expanding the singular parts
of the resolvents as follows
\beq
g_1(q)=g_1^{(0)}(q)+\frac{1}{J}\,g_1^{(1)}(q),
\hspace{0.7cm}g_2(q)=g_2^{(0)}(q)+\frac{1}{J}\,g_2^{(1)}(q).
\eeq

\section{Particular points \label{particular}}
Above we have derived the conformal dimension
including the non-anomalous part of the
finite size correction
for all rational three-spin strings by a one line computation.
One word of caution is needed, though.
Our expression for $\gamma$ constitutes  a 
necessary condition
for the Bethe equations to be fulfilled. Implicitly we have assumed that no
root of the first type coincides with a root of the second type. We need
to check that the full solution of the Bethe equations indeed has this  
property.
In the thermodynamic limit, $J\rightarrow \infty$, the Bethe roots
condense on smooth contours in the complex plane.
Thus we should investigate whether the contours corresponding to
$\{q_{1,j}\}_{j=1}^{n_1}$ and
$\{q_{2,j}\}_{j=1}^{n_2}$ are indeed disjunct. The supports for the distributions of roots constitute the 
branch cuts of the resolvents $G_1(q)$ and $G_2(q)$. 
To determine the branch points of the resolvents we only need to consider the
$J\rightarrow\infty$ limit of eqns.~\rf{quadratic} and~\rf{cubic}.
The branch points of the resolvents are thus given by the single zeroes of
the common discriminant of the two 
cubic equations~\rf{g1cubic} and~\rf{g2cubic}. The discriminant 
reads
\beq
\Delta=4 (r(q))^3- 27(s(q))^2 \label{delta}.
\eeq
Inserting eqns.~\rf{rqexp} and~\rf{sqexp} in eqn.~\rf{delta} we find that the 
leading and next to leading order term in $\frac{1}{q}$ cancel out so that
$\Delta$ becomes a polynomial of degree four in $\frac{1}{q}$. The
four zeroes of this polynomial then constitute the branch points for
$G_1(q)$ and $G_2(q)$.
In general one can not determine the precise location of the cuts from the
position of the branch points alone.
In order that a meaningful density of Bethe roots can be associated
with the resolvents the discontinuity of these across their respective
cuts, i.e.\
$\rho(q)=\frac{1}{2\pi i}\left\{G(q+i0)-G(q-i0)\right\}$,
 must fulfill that
$\rho(q) dq$ is real and positive~\cite{David:1990sk}.

There are in general three directions consistent with positivity from which
a cut can emerge from a branch point~\cite{David:1990sk}. 
We expect that in general it will be
possible to choose the cuts so that they do not overlap. 
(For an example of how this works, see~\cite{Ambjorn:1994bp}).
 There is, however, one situation where we can
detect a signal of our solution coming to a limitation, namely the situation
where two or more of the four branch points coincide. This either means that
one (or two) cuts degenerate to a point or that two cuts touch
each other.
As the general expression for $\Delta$ is rather involved, 
let us specialize to the cases where two spin labels coincide.

Let us consider the case $\beta=\frac{\alpha}{2}$, $n=-2m=k$
corresponding to a string with spin assignment $(J_1,J_2,J_2)$,
$J_1>J_2$ and winding number $k$. This is the case which includes the
stable three-spin string.
Here, the allowed region for $\alpha$ is $0\leq\alpha\leq\frac{2}{3}$.
In this case we get for the discriminant
\beq
\Delta \propto (1-\alpha)-
q^2(8-36\alpha+27\alpha^2)(k \pi)^2+q^4(2\pi k)^4.
\eeq
This polynomial has  two sets of double roots at $q=\pm i \sqrt{3} (2 \pi k)$
when 
\beq
\alpha=\frac{8}{9}.
\eeq
This value of $\alpha$ was likewise singled out in the analysis 
of~\cite{Engquist:2003rn} for reasons similar to the one described here.
We notice that the particular point
$\alpha=\frac{8}{9}$ lies outside the allowed region (and thus does
not affect our comparison with string theory).
Exploiting the symmetry of the equation~\rf{s1} and~\rf{s2} under 
$\frac{\alpha}{2}\rightarrow 1-\alpha$ we immediately find that for a
string with spin assignment $(J_1,J_1,J_3)$ with $J_1>J_3$ the
corresponding
particular value of $\alpha$ is
\beq
\alpha=\frac{5}{9}.
\eeq
This value of $\alpha$ was likewise singled out in the analysis
of~\cite{Engquist:2003rn}. We expect to have an entire line of
particular points $\alpha=\alpha_{*}(\beta)$ in the parameter space.

\section{Conclusion \label{conclusion}}

We have determined the one-loop conformal dimension and
the
associated non-anomalous
finite size corrections for all operators dual to rational
three-spin strings. The conformal dimensions match the classical
string energies and for operators dual to stable three-spin strings
with two coinciding angular momenta the non-anomalous part of the 
finite size correction agrees with the zero-mode contribution to the 
one-loop string sigma model energy.
 Very recently there was another comparison
of gauge theory non-anomalous
finite size corrections and zero-mode contributions to string one-loop
energies which likewise resulted in agreement~\cite{Park:2005ji}. 
This comparison
concerned a stable two-spin string with one large angular momentum on
$S^5$ and one on $AdS_5$. We expect that it should be relatively 
straightforward to extend our analysis of the three-spin case to the situation
where one or two of the spins lie in $AdS_5$ in stead of $S^5$. It
appears less straightforward to treat the case of elliptic or hyper-elliptic
strings. Rather than extending the analysis to further particular solutions
it would of course be more interesting to find a unifying geometric
description of the finite size corrections applicable to any string solution 
--- in the spirit of the treatment of the leading order
contribution~\cite{Kazakov:2004qf,Kazakov:2004nh,Beisert:2004ag,Schafer-Nameki:2004ik}. 

As mentioned in the introduction, 
it has been known for some time that to leading order in $\frac{1}{J}$
there
exists a discrepancy between the classical energy of spinning strings
and
the conformal dimension of the dual operators at third order in
$\frac{\lambda}{J^2}$~\cite{Serban:2004jf}, see
also~\cite{Minahan:2004ds}.
For near BMN states a similar three-loop disagreement is
observed between gauge- and string theory at next to leading order
in $\frac{1}{J}$~\cite{Beisert:2003tq,Callan:2003xr}.
Therefore, for spinning strings at next to leading
order in $\frac{1}{J}$    
 one would likewise expect a disagreement to turn up 
at some low order in
$\frac{\lambda}{J^2}$. Our results as well as the results 
of~\cite{Park:2005ji} indicate that this does not happen at 
first order in $\frac{\lambda}{J^2}$ and the forthcoming
paper~\cite{Beisert:2005mq}, which takes into account the anomalous
part of the finite size correction, find complete agreement 
at this order for the
case of a two-spin string with one large angular momentum on $S^5$ and
one on $AdS_5$. It would of course be interesting to determine the
anomalous part of the finite size correction for the three-spin
strings as well. This would require the explicit knowledge of the
leading order Bethe root distributions, an information which
is in principle accessible but in practice hard to obtain, cf.
section~\ref{particular}.
In any
case, we have with our investigations enlarged the region of parameter
space  where a comparison of semi-classical strings and their dual
operators is possible and have as well provided new data which might
help in further refining the methods for comparing gauge- and string
theory.

\vspace*{1.0cm}
\noindent
{\bf Acknowledgements}

We thank A.\ Luther, J.\ Minahan, T.\ M\aa nsson, 
M.\ Smedb$\ddot{a}$ck, M. Staudacher and \mbox{O.\ Sylju\aa sen}
for discussions. Furthermore, we are grateful to N.\ Beisert, A.\
Tseytlin and K.\ Zarembo for sharing with us their
manuscript~\cite{Beisert:2005mq} prior to
publication.


\vspace*{0.5cm}


\begin{thebibliography}{999}

\bibitem{Tseytlin:2003ii}
A.~A.~Tseytlin,
hep-th/0311139.

\bibitem{Maldacena:1997re}
J.~M.~Maldacena,
Adv.\ Theor.\ Math.\ Phys.\  {\bf 2} (1998) 231
[Int.\ J.\ Theor.\ Phys.\  {\bf 38} (1999) 1113], hep-th/9711200;
S.~S.~Gubser, I.~R.~Klebanov and A.~M.~Polyakov,
Phys.\ Lett.\ B {\bf 428} (1998) 105, hep-th/9802109;
E.~Witten,
Adv.\ Theor.\ Math.\ Phys.\  {\bf 2} (1998) 253, hep-th/9802150.

\bibitem{Frolov:2003qc}
S.~Frolov and A.~A.~Tseytlin,
Nucl.\ Phys.\ B {\bf 668} (2003) 77,
hep-th/0304255.

\bibitem{Frolov:2003tu}
S.~Frolov and A.~A.~Tseytlin,
JHEP {\bf 0307} (2003) 016, hep-th/0306130.

        
\bibitem{Minahan:2002ve}
J.~A.~Minahan and K.~Zarembo,
JHEP {\bf 0303}, 013 (2003),
hep-th/0212208.
   
\bibitem{Beisert:2003jj}
N.~Beisert,
Nucl.\ Phys.\ B {\bf 676} (2004) 3,
hep-th/0307015.

\bibitem{Beisert:2003yb}
N.~Beisert and M.~Staudacher,
Nucl.\ Phys.\ B {\bf 670} (2003) 439,
hep-th/0307042.

\bibitem{Beisert:2003xu}
N.~Beisert, J.~A.~Minahan, M.~Staudacher and K.~Zarembo,
JHEP {\bf 0309} (2003) 010,
hep-th/0306139;

\bibitem{Beisert:2003ea}
N.~Beisert, S.~Frolov, M.~Staudacher and A.~A.~Tseytlin,
JHEP {\bf 0310} (2003) 037,
hep-th/0308117.

\bibitem{Engquist:2003rn}
J.~Engquist, J.~A.~Minahan and K.~Zarembo, JHEP {\bf 0311} (2003) 063,
hep-th/0310188.

\bibitem{Kristjansen:2004ei}
C.~Kristjansen,
Phys.\ Lett.\ B {\bf 586} (2004) 106,
hep-th/0402033.

\bibitem{Kristjansen:2004za}
C.~Kristjansen and T.~Mansson,
Phys.\ Lett.\ B {\bf 596}, 265 (2004),
hep-th/0406176.



\bibitem{Kruczenski:2003gt}
M.~Kruczenski,
Phys.\ Rev.\ Lett.\  {\bf 93} (2004) 161602, hep-th/0311203.



\bibitem{Kazakov:2004qf}
V.~A.~Kazakov, A.~Marshakov, J.~A.~Minahan and K.~Zarembo,
JHEP {\bf 0405} (2004) 024, hep-th/0402207;

\bibitem{Hernandez:2004uw}
R.~Hernandez and E.~Lopez,
JHEP {\bf 0404} (2004) 052
hep-th/0403139.



\bibitem{Kruczenski:2004kw}
M.~Kruczenski, A.~V.~Ryzhov and A.~A.~Tseytlin,
Nucl.\ Phys.\ B {\bf 692} (2004) 3, hep-th/0403120.



\bibitem{Stefanski:2004cw}
B.~J.~Stefanski and A.~A.~Tseytlin,
JHEP {\bf 0405} (2004) 042, hep-th/0404133.

\bibitem{Hernandez:2004kr}
R.~Hernandez and E.~Lopez,
JHEP {\bf 0411}, 079 (2004),
hep-th/0410022.

\bibitem{Kazakov:2004nh}
V.~A.~Kazakov and K.~Zarembo,
JHEP {\bf 0410} (2004) 060,
hep-th/0410105.

\bibitem{Beisert:2004ag}
N.~Beisert, V.~A.~Kazakov and K.~Sakai,
hep-th/0410253.

\bibitem{Schafer-Nameki:2004ik}
 S.~Schafer-Nameki,
 hep-th/0412254.

\bibitem{Beisert:2003tq}
N.~Beisert, C.~Kristjansen and M.~Staudacher,
Nucl.\ Phys.\ B {\bf 664} (2003) 131, hep-th/0303060.

\bibitem{Beisert:2003ys}
N.~Beisert, Nucl.\ Phys.\ B {\bf 682} (2004) 487,
hep-th/0310252.

\bibitem{Staudacher:2004tk}
M.~Staudacher,
hep-th/0412188.


\bibitem{Serban:2004jf}
D.~Serban and M.~Staudacher,
JHEP {\bf 0406} (2004) 001,
hep-th/0401057.

\bibitem{Minahan:2004ds}
J.~A.~Minahan,
JHEP {\bf 0410}, 053 (2004),
hep-th/0405243.

\bibitem{Beisert:2004hm}
N.~Beisert, V.~Dippel and M.~Staudacher,
JHEP {\bf 0407} (2004) 075,
hep-th/0405001.


\bibitem{Peeters:2004pt}
K.~Peeters, J.~Plefka and M.~Zamaklar,
JHEP {\bf 0411} (2004) 054
hep-th/0410275.


\bibitem{Frolov:2004bh}
S.~A.~Frolov, I.~Y.~Park and A.~A.~Tseytlin,
Phys.\ Rev.\ D {\bf 71} (2005) 026006, hep-th/0408187.


\bibitem{Lubcke:2004dg}
M.~Lubcke and K.~Zarembo,
JHEP {\bf 0405}, 049 (2004),
hep-th/0405055.


\bibitem{Park:2005ji}
I.~Y.~Park, A.~Tirziu and A.~A.~Tseytlin,
hep-th/0501203.

\bibitem{Beisert:2005mq}
N.~Beisert, A.~A.~Tseytlin and K.~Zarembo,
Nucl.\ Phys.\ B {\bf 715} (2005) 190
hep-th/0502173.

\bibitem{Arutyunov:2003za}
G.~Arutyunov, J.~Russo and A.~A.~Tseytlin,
Phys.\ Rev.\ D {\bf 69} (2004) 086009,
hep-th/0311004.

\bibitem{Arutyunov:2003uj}
G.~Arutyunov, S.~Frolov, J.~Russo and A.~A.~Tseytlin,
Nucl.\ Phys.\ B {\bf 671} (2003) 3
hep-th/0307191.

\bibitem{Bethe:1931hc}
H.~Bethe,
Z.\ Phys.\  {\bf 71} (1931) 205.

\bibitem{Hailu:2004dq}
G.~Hailu,
JHEP {\bf 0502}, 017 (2005), hep-th/0411256.


\bibitem{Freyhult:2004iq}
L.~Freyhult,
JHEP {\bf 0406} (2004) 010,
hep-th/0405167.


\bibitem{Arutyunov:2003rg}
G.~Arutyunov and M.~Staudacher,
JHEP {\bf 0403} (2004) 004,
hep-th/0310182;
G.~Arutyunov and M.~Staudacher,
hep-th/0403077.


\bibitem{Engquist:2004bx}
J.~Engquist,
JHEP {\bf 0404} (2004) 002,
hep-th/0402092.



\bibitem{Kostov:1988fy}
I.~K.~Kostov,
Mod.\ Phys.\ Lett.\ A {\bf 4}, 217 (1989);
I.~K.~Kostov and M.~Staudacher,
Nucl.\ Phys.\ B {\bf 384}, 459 (1992),
hep-th/9203030;
B.~Eynard and J.~Zinn-Justin,
Nucl.\ Phys.\ B {\bf 386} (1992) 558,
hep-th/9204082;
B.~Eynard and C.~Kristjansen,
Nucl.\ Phys.\ B {\bf 455}, 577 (1995),
hep-th/9506193.


\bibitem{David:1990sk}
F.~David,
Nucl.\ Phys.\ B {\bf 348}, 507 (1991).

\bibitem{Ambjorn:1994bp}
J.~Ambj\o rn, C.~F.~Kristjansen and Y.~Makeenko,
Phys.\ Rev.\ D {\bf 50}, 5193 (1994),
hep-th/9403024.




\bibitem{Callan:2003xr}
C.~G.~.~Callan, H.~K.~Lee, T.~McLoughlin, J.~H.~Schwarz, I.~Swanson and X.~Wu,
Nucl.\ Phys.\ B {\bf 673} (2003) 3,
hep-th/0307032;
C.~G.~.~Callan, T.~McLoughlin and I.~Swanson,
Nucl.\ Phys.\ B {\bf 694} (2004) 115,
hep-th/0404007;
C.~G.~.~Callan, T.~McLoughlin and I.~Swanson,
Nucl.\ Phys.\ B {\bf 700} (2004) 271,
hep-th/0405153.


\end{thebibliography}
\end{document}